\documentclass[journal]{IEEEtran}
\usepackage{amsmath,graphicx,url}
\usepackage[caption=false,font=footnotesize,skip=1pt,subrefformat=parens]{subfig}
\usepackage[hidelinks]{hyperref}
\newcommand{\hex}[1]{\texttt{0x#1}}
\begin{document}
\bstctlcite{IEEEexample:BSTcontrol}
\title{Decoding Broadcast Control Information in the Starlink Ku-Band Downlink}
\author{Haoyan Liu, Yichuan Ma, Haotong Zhao, Chengguang Li, and Min Yang%
\thanks{The authors are with the School of Aerospace Science and Technology, Xidian University, Xi'an 710071, China.}}
\maketitle
\begin{abstract}
This correspondence analyzes the control messages carried by the Starlink Ku-band downlink allocation using modulation and coding scheme (MCS) index~5. Across 1009 publicly available frames, an inner header protected by a 24-bit cyclic redundancy check (CRC) identifies the boundaries of 3722 messages and specifies the trailing padding. Consistency checks based on satellite motion and timestamp progression support the interpretation of fields encoding satellite position, velocity, and timing. The leading message also contains a fixed ground reference point. Two message types share a resource-list format, and the parsed list lengths agree with the inner-header length descriptors for all 1355 list-bearing messages. Each 60-bit descriptor encodes a labeled two-dimensional resource allocation. The observed grouping and partitioning are consistent with the uplink resource map described in a SpaceX patent. The physical units of the resource axes and the mapping of labels to terminals remain unresolved.
\end{abstract}
\section{Introduction}
Passive measurements have established the frame timing, synchronization sequences, and orthogonal frequency-division multiplexing (OFDM) parameters of the Starlink Ku-band downlink \cite{humphreys2023}. Subsequent studies characterized frame arrival times, identified pilots and other predictable elements, and released equalized frames \cite{qintiming,qin2026,qindata}. Similar passive measurements have also yielded waveform parameters and demodulated symbols for OneWeb \cite{oneweb}. These studies provide the basis for physical-layer signal recovery. However, identifying the control messages carried by the decoded bytes requires additional evidence to establish their boundaries and interpret their field semantics.

In every analyzed frame, the allocation between the first and second protocol data unit (PDU) headers uses modulation and coding scheme (MCS) index~5, with binary phase-shift keying (BPSK) and rate-$1/2$ low-density parity-check (LDPC) coding \cite{prior}. Its recurring position and robust transmission motivate an investigation into whether it carries broadcast control information. SpaceX patents describe the broadcast of initial-access information and the delivery of uplink assignments over the downlink \cite{entry,scheduler}. However, the example MCS table in \cite{mcs} does not include index~5, leaving its role in this allocation to be established through the decoded message contents.

Using 1009 public frames from STARLINK-31848, this correspondence recovers a cyclic redundancy check (CRC)-protected inner header and validates the message boundaries against the decoded allocation length. Kinematic and timing consistency checks support the interpretation of satellite-state fields in the leading message, which also contains a fixed ground reference. Two other message types share a list format for describing labeled resource intervals. The observed grouping and partitioning of these intervals, together with their correspondence to the patented uplink resource map, support a candidate uplink-grant interpretation \cite{scheduler}.

\section{Inner Header and Message Boundaries}
\textit{Source-bit recovery.}
The dataset provided by the University of Texas at Austin contains soft frequency-domain symbols obtained after timing and phase correction as well as power normalization \cite{qin2026,qindata}. For each OFDM symbol, the receiver serializes the active non-pilot subcarriers and multiplies the resulting symbols by the complex conjugate of a fixed four-phase reference sequence. The first protocol data unit (PDU) header is then recovered through convolutional decoding and validated using CRC-8 \cite{prior}. In the public frames, the allocation immediately following this header is signaled with MCS index~5. Its physical-layer length follows from the codeword count \cite{waveform}. The allocation uses binary phase-shift keying (BPSK) and rate-$1/2$ low-density parity-check (LDPC) coding with 2048-bit codewords \cite{prior}. Each BPSK data resource element carries one coded bit. The receiver selects the constellation orientation that minimizes the number of failed parity checks over three public frames. The resulting soft bit metrics feed a normalized min-sum decoder with the fixed parity-check matrix reported in \cite{prior}. The 1009 frames yield 4062 decoded LDPC codewords, each with 1024 systematic bits.

Additive descrambling operates on the decoded systematic bits after bit reversal within each octet. The sequence $s[n]$ is initialized for each codeword with the 15-bit representation of \hex{6FC0} in least-significant-bit-first order and follows $s[n]=s[n-14]\oplus s[n-15]$ for $n\ge15$, where $\oplus$ denotes exclusive OR (XOR). The descrambled bits are packed most significant bit first. The resulting 128-byte blocks are concatenated in allocation order.

\begin{figure}[!t]
\centering
\subfloat[]{\includegraphics[width=0.96\columnwidth]{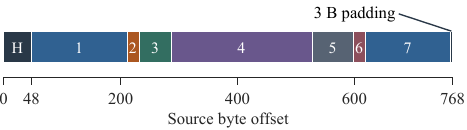}\label{fig:mux:allocation}}\par
\subfloat[]{\includegraphics[width=0.96\columnwidth]{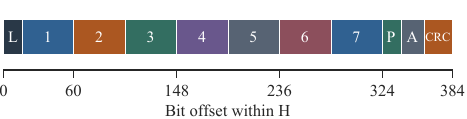}\label{fig:mux:header}}
\caption{Decoded MCS-5 allocation in F0001. (a) The 768-byte allocation begins with the 48-byte inner header H. SDUs 1--7 have types \hex{00}, \hex{08}, \hex{02}, \hex{01}, \hex{03}, \hex{03}, and \hex{03}, and lengths 163, 21, 55, 241, 71, 20, and 146 bytes. (b) Expanded bit layout of H. L is the header length, 1--7 are the SDU length descriptors, P specifies allocation padding, and A denotes header alignment bits.}
\label{fig:mux}
\end{figure}

\textit{Structure of the decoded allocation.}
The decoded MCS-5 allocation begins with an inner header that provides the length information for the subsequent service data units (SDUs) and indicates the amount of trailing zero-byte padding. Fig.~\ref{fig:mux}\subref{fig:mux:allocation} illustrates the arrangement of the header, SDUs, and padding within the allocation. The first two bytes encode the header length in bytes, while the final three bytes contain the CRC-24. Once the header CRC is validated, the length descriptors are used sequentially to determine the boundaries of the SDUs. Each SDU begins with a type byte, followed by a 3-bit field indicating the number of trailing zero bits. A related length-delimiting mechanism is used in Long Term Evolution (LTE) Radio Link Control (RLC) data PDUs, whose header extension may contain multiple length indicators (LIs) preceding the data field, each specifying the length of a corresponding data-field element \cite{lterlc}.

\textit{Inner header and length validation.}
Fig.~\ref{fig:mux}\subref{fig:mux:header} expands the inner header. Table~\ref{tab:header} gives its field offsets. Bit indices within each byte start at the least significant bit, and bit $j$ of an unsigned field has weight $2^j$. The first 16 bits specify the header length $L$ in bytes. The CRC occupies its last 24 bits. In the standard polynomial representation \cite{koopman}, the CRC generator has degree 24 and lower coefficients given by \hex{1F21C5}. The register starts at \hex{0AE0BC}. The XOR of each protected bit and the most significant register bit gives the feedback. The register shifts left within 24 bits and is XORed with \hex{1F21C5} when the feedback is one. The final XOR is \hex{F51F43}. Input and check bytes are read least significant bit first. The first check bit occupies the most significant position of the comparison word.

Each 44-bit descriptor contains the observed context pattern \hex{0FF001}, tag~6, and the SDU length $\ell_j$ in bytes. Tag~8 terminates the list of $J$ descriptors and introduces the 12-bit allocation-padding length $P$. With one-valued alignment bits and the CRC included, the header occupies $L=4\lceil(56+44J)/32\rceil$ bytes. The sum of the header, SDU, and padding lengths must also agree with the source-byte capacity of the $M$ LDPC codewords signaled by the physical-layer header
\begin{equation}
L+\sum_{j=0}^{J-1}\ell_j+P=128M.
\end{equation}

\begin{table}[!t]
\caption{Inferred Structure of the Inner Header}
\label{tab:header}
\centering
\footnotesize
\begin{tabular}{@{}llp{0.51\columnwidth}@{}}
\hline
Bit start & Width & Field and observed value\\
\hline
$0$ & $16$ & Header length $L$ in bytes\\
$h_j$ & $24$ & Context pattern \hex{0FF001}\\
$h_j+24$ & $4$ & Length-descriptor tag 6\\
$h_j+28$ & $16$ & Length $\ell_j$ of SDU $j$ in bytes\\
$B$ & $4$ & End-of-list tag 8\\
$B+4$ & $12$ & Source padding $P$ in bytes\\
$B+16$ & Variable & Ones through bit $8L-25$\\
$8L-24$ & $24$ & Header CRC\\
\hline
\end{tabular}
\parbox{0.92\columnwidth}{\vspace{3pt}$h_j=16+44j$ for $0\le j<J$ and $B=16+44J$. An empty alignment interval is allowed.}
\end{table}
Frame F0001 illustrates both checks. In the header shown in Fig.~\ref{fig:mux}\subref{fig:mux:header}, the bytes \texttt{30 00} give $L=48$, and the stored CRC bytes \texttt{E8 05 81} assemble to \hex{17A081}. Seven descriptors give $B=324$ and 20 alignment bits. The first length at bit~44 is 163 bytes, placing the second SDU at byte 211. The allocation in Fig.~\ref{fig:mux}\subref{fig:mux:allocation} totals $48+717+3=768$ bytes, the source capacity of six codewords.

All 1009 headers passed CRC-24, the header-structure length check, and the allocation-capacity check. Header lengths ranged from 20 to 60 bytes. The descriptors delimited 3722 SDUs of seven types, and reassembling the intervals and padding reproduced each decoded allocation.

\section{Satellite State, Timing, and Optional Records}
\subsection{State Vector and Timing Fields}
The first SDU in every analyzed frame has type \hex{00}. The use of ephemeris information during terminal discovery, described in the network-entry patent \cite{entry}, motivates testing for satellite state fields. The 16-bit value at bit~11 matched the dataset satellite number 31848 in all frames. The full identifier format and namespace remain unresolved. Position components start at bits 60, 124, and 188 in binary64 format. Velocity components start at bits 252, 284, and 316 in binary32 format. The 32-bit unsigned values at bits 348 and 380 are interpreted as the seconds and nanosecond components of a timestamp.

\begin{table}[!t]
\caption{Inferred Fields of the Type-\hex{00} Message}
\label{tab:statefields}
\centering
\footnotesize
\begin{tabular}{@{}llp{0.49\columnwidth}@{}}
\hline
Bit start & Width & Field and reading convention\\
\hline
0, 8 & 8, 3 & Type and terminal zero-bit count\\
11, 27 & 16, 33 & Candidate satellite label and uninterpreted bits\\
60, 124, 188 & $3\times64$ & XYZ position, binary64\\
252, 284, 316 & $3\times32$ & XYZ velocity, binary32\\
348, 380 & $32,32$ & Candidate timestamp seconds and nanoseconds\\
412, 413 & $1,8$ & Uninterpreted bit and 8-bit value repeated in messages \hex{01} and \hex{03}\\
421, 444 & $23,10$ & Coarse timing field $C$ and uninterpreted bits\\
454, 472 & $18,48$ & Fine timing field $F$ and fixed bytes\\
520 & 224 & Four 56-bit records in long layout\\
$8G-80$ & 80 & Fixed preamble to ground coordinates\\
$8G+64i$ & $3\times64$ & Ground XYZ, binary64, $i=0,1,2$\\
$8G+192$ & 288 & Remaining bytes and terminal padding\\
\hline
\end{tabular}
\parbox{0.92\columnwidth}{\vspace{3pt}$G=75$ or 103 is the byte offset of ground XYZ in the short or long layout, respectively. Widths with a list of starts are per field.}
\end{table}

Table~\ref{tab:statefields} summarizes the inferred fields, including the Institute of Electrical and Electronics Engineers (IEEE) 754 position and velocity components \cite{ieeefloat}. For F0001, these decode to $(-496.287,-5787.423,3569.847)$\,km and $(6296.41,1525.49,3339.75)$\,m/s. The timestamp fields contain $1421546977$ and $389157118$. The meaning of the 33 bits immediately following the satellite identification field has yet to be determined. Further examination of the remaining fields reveals that the value of the 8-bit field starting at bit~413 is also observed in the subsequent messages \hex{01} and \hex{03}, while the six bytes starting at bit~472 form the hexadecimal sequence \texttt{06 A0 76 C5 2E 15}.

\begin{figure}[!t]
\centering
\subfloat[]{\includegraphics[width=0.85\columnwidth]{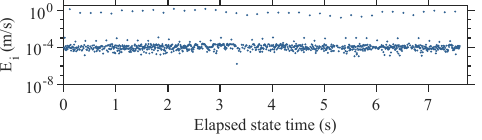}\label{fig:state:kinematic}}\par
\subfloat[]{\includegraphics[width=0.85\columnwidth]{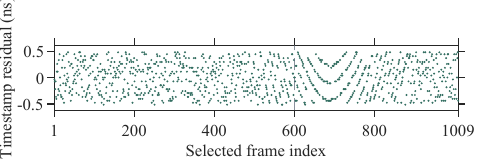}\label{fig:state:timing}}
\caption{Consistency of the decoded satellite state and timing fields. (a) Kinematic consistency error for the 992 intervals with positive timestamp increments. (b) Timestamp residual from \eqref{eq:clock}, with the constant offset estimated from the 600 frames to the left of the dashed line.}
\label{fig:state}
\end{figure}

The state-vector interpretation can be tested by comparing the velocity inferred from successive positions with the decoded velocities. Let $\boldsymbol r_i$, $\boldsymbol v_i$, and $t_i$ denote the decoded position, velocity, and timestamp of selected frame $i$. For $t_{i+1}>t_i$, the kinematic consistency error is
\begin{equation}
E_i=\left\|
\frac{\boldsymbol r_{i+1}-\boldsymbol r_i}{t_{i+1}-t_i}
-\frac{\boldsymbol v_{i+1}+\boldsymbol v_i}{2}\right\|_2.
\end{equation}
Of 1008 adjacent intervals, 992 had positive timestamp increments and 16 repeated both timestamp and position. Fig.~\ref{fig:state}\subref{fig:state:kinematic} shows the errors for those 992 intervals, with a median of $1.03\times10^{-4}$\,m/s and a maximum of 1.39\,m/s. Shifting all three position offsets by $-2$, $-1$, $+1$, or $+2$ bits gave no interval with $E_i$ below 2\,m/s. This supports the inferred offsets and interpretation of the position and velocity fields.

Let $C$ and $F$ denote the 23-bit coarse field at bit~421 and the 18-bit fine field at bit~454. Relative to F0001, the decoded timestamp increments are consistent with
\begin{equation}
\Delta t=\Delta C/750-\Delta F/(270\times10^6)\quad\text{s}.
\label{eq:clock}
\end{equation}
A unit increment in $C$ adds about 1.333\,ms, and a unit increment in $F$ subtracts about 3.704\,ns. Between F0001 and F0002, $C$ increases from 5540 to 5544 and $F$ from 114514 to 114540. Equation~\eqref{eq:clock} predicts 5,333,237.037\,ns, compared with the decoded increment of 5,333,237\,ns.

The first 600 frames set a constant residual offset. After applying it, the maximum absolute residual over the other 409 frames was 0.513\,ns (Fig.~\ref{fig:state}\subref{fig:state:timing}). Moving the fine-field start to bit~453 or 455, or shortening it to 17 bits at bit~454, increased the maximum to 934, 39.6, or 483\,$\mu$s, respectively. These residuals support the field boundaries and their relative timing relationship. The absolute time reference and accuracy relative to an external clock remain unresolved.

\subsection{Ground Reference Coordinates}
\begin{figure}[!t]
\centering
\includegraphics[width=0.96\columnwidth]{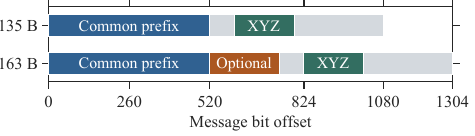}
\caption{Ground-coordinate placement in the two type-\hex{00} layouts. The ground XYZ field starts at bit~600 in the 135-byte layout. The optional 28-byte block shifts it to bit~824 in the 163-byte layout.}
\label{fig:groundlayout}
\end{figure}

As illustrated in Fig.~\ref{fig:groundlayout}, two structural variants of the type-\hex{00} SDU are observed, with lengths of 135 and 163 bytes, respectively. These variants differ in the presence of an optional 28-byte block starting at bit~520, which shifts the starting bit index of the ground coordinate field from bit~600 in the short layout to bit~824 in the long layout. In both layouts, the three coordinate components represented in binary64 format are immediately preceded by the same 10-byte sequence, \texttt{80 01 4A D6 9E 59 00 00 0A 80}. The subsequent 36 bytes begin with \hex{6B} and contain fields whose functions have yet to be determined. Frames F0001 and F0002 provide representative examples of the long and short layouts, respectively.

The decoded coordinate vector is approximately $(-731363.1,$ $-5463857.1,$ $3197359.3)$\,m. When interpreted in the World Geodetic System 1984 (WGS~84) Earth-centered, Earth-fixed (ECEF) coordinate system \cite{wgs84}, this vector corresponds to a latitude of $30.28149^\circ$\,N and a longitude of $97.62399^\circ$\,W, with an ellipsoidal height of approximately zero. The decoded coordinates place the reference point in the Austin, Texas, area. The University of Texas at Austin also reports receiving STARLINK-31848 in Austin \cite{austincampaign}. Agreement at the city scale provides a geographic check on the decoded ground coordinates and supports their interpretation as a ground reference. The reference point's operational function remains unknown.

\subsection{Structure of the Optional Block}
\begin{figure}[!t]
\centering
\includegraphics[width=0.96\columnwidth]{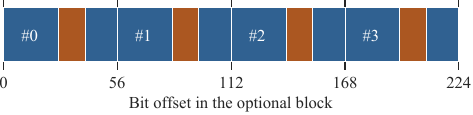}
\caption{The optional 28-byte block in the long type-\hex{00} layout. Each of the four 56-bit records contains a copy of the same 13-bit field, shown in orange.}
\label{fig:groundopts}
\end{figure}

Fig.~\ref{fig:groundopts} illustrates the internal structure of the optional 28-byte block, which consists of four 56-bit records indexed from 0 to 3. The horizontal axis represents the bit offset relative to the beginning of the block at bit~520 of the SDU. Accordingly, record $j$ starts at bit $520+56j$ within the SDU, where $j=0,1,2,3$. The detailed field structure of each record is summarized in Table~\ref{tab:optionfields}. The two 2-bit fields starting at record-relative bit indices 23 and 45 both correspond to the record index $j$. Of particular interest is the 13-bit field spanning record-relative bits 27--39. This field remains invariant across the four records within each SDU but varies between SDUs. For example, the first record in frame F0001 is \texttt{01 40 3F 00 11 00 24}, and all four records in this frame share a field value of 544. Across the 718 long messages, the value of this field ranges from 542 to 583.

\begin{table}[!t]
\caption{Structure of a Seven-Byte Record in the Optional Block}
\label{tab:optionfields}
\centering
\footnotesize
\begin{tabular}{@{}rrl@{}}
\hline
Bit start & Width & Value in records $j=0,1,2,3$\\
\hline
0 & 1 & One only for $j=0$\\
1, 25, 40, 47 & --- & Zero windows of 7, 2, 5, and 1 bits\\
8 & 8 & \hex{40}\\
16 & 7 & 63\\
23, 45 & $2$ each & Record ordinal $j$\\
27 & 13 & Common value across the four records\\
48 & 8 & \hex{24}\\
\hline
\end{tabular}
\end{table}

\section{Resource Allocation Messages and Candidate Uplink Grants}
\subsection{Resource-List Location and Boundaries}

\begin{figure}[!t]
\centering
\includegraphics[width=0.98\columnwidth]{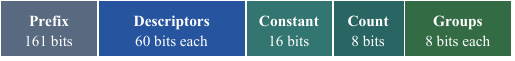}
\caption{Layout of a list-bearing SDU before byte alignment. The count gives the number of group descriptors in the final field. Block widths are schematic.}
\label{fig:resource-list-layout}
\end{figure}

Based on the inner-header length fields recovered in Section~II, we identify 874 type-\hex{01} SDUs and 555 type-\hex{03} SDUs. Both SDU types share the common message structure summarized in Fig.~\ref{fig:resource-list-layout} and Table~\ref{tab:listfields}. All bit offsets are defined relative to the beginning of each SDU, with indexing starting at zero and following the least-significant-bit-first convention described in Section~II.

\begin{table}[!t]
\caption{Observed Layouts of Type-\hex{01} and Type-\hex{03} Messages}
\label{tab:listfields}
\centering
\footnotesize
\begin{tabular}{@{}llp{0.53\columnwidth}@{}}
\hline
Bit start & Width & Field or observed rule\\
\hline
0, 8 & 8, 3 & Message type and terminal zero-bit count\\
11 & 138 & Uninterpreted prefix bits\\
149 & 4 & List tag, equal to 0 or 7\\
153 & 8 & Record count \(N\) when the tag is 7\\
161 & \(60N\) & Sequence of \(N\) resource descriptors\\
\(D\), \(D+16\) & 16, 8 & Fixed value \hex{0100} and group count \(K\)\\
\(D+24\) & \(8K\) & \(K\) group descriptors, followed by zero alignment\\
\hline
\end{tabular}
\parbox{0.92\columnwidth}{\vspace{3pt}For the list-bearing branch, \(D=161+60N\). When the tag is 0, bits 153--159 are zero and the message ends at bit~160, exclusive.}
\end{table}

The 4-bit tag starting at bit~149 distinguishes between two message layouts. When the tag is set to~0, the SDU has a fixed length of 20 bytes and contains no resource list. When the tag is set to~7, the SDU contains a sequence of \(N\) consecutive 60-bit resource descriptors, where \(N\) is specified by the 8-bit field starting at bit~153. These descriptors are tentatively interpreted as uplink grants and collectively occupy the bit interval \([161,161+60N)\). Among the 1429 analyzed SDUs, 74 follow the former layout and 1355 follow the latter.

Immediately following the resource descriptors, the list footer starts at bit \(D=161+60N\) and consists of a 16-bit field with a fixed value of \hex{0100}, an 8-bit group count \(K\), and \(K\) one-byte group descriptors. Based on the inferred message structure, the total SDU length, including zero padding for byte alignment, can be expressed as
\begin{equation}
L_{\mathrm{SDU}}=
\left\lceil\frac{185+60N+8K}{8}\right\rceil
\label{eq:sdu_length}
\end{equation}
in bytes. For all 1355 SDUs containing resource lists, the calculated lengths agree exactly with those indicated by the corresponding CRC-validated inner headers, providing quantitative evidence for the inferred message structure and field boundaries.

Taking frame F0001 as an example, a type-\hex{01} SDU begins at byte~287 of the source byte stream and contains \(N=28\) resource descriptors, which occupy bits \([161,1841)\), followed by the list footer starting at bit~1841. With a group count of \(K=7\), the SDU comprises 1921 bits before byte alignment and occupies 241 bytes after padding. Accordingly, the next SDU begins at byte~528 of the source byte stream. The same frame also contains a type-\hex{03} SDU with a tag value of~0 and a length of 20 bytes.

\subsection{Decoding the Resource Descriptors}

Table~\ref{tab:resource} illustrates the internal structure of a resource descriptor, which contains a 12-bit label provisionally interpreted as a service identifier (SID), a group index, and four parameters specifying a two-dimensional resource allocation. The resource intervals along the two axes are given by \([a,a+d)\) and \([b,b+e+1)\), where \(a\) and \(b\) denote the respective starting indices. Unlike the first-axis allocation length, which is directly represented by \(d\), the second-axis allocation length is encoded as \(e+1\). A comparable length-minus-one representation is adopted in the resource indication value (RIV) for fifth-generation New Radio (5G NR) Type-1 downlink resource allocation \cite{nrphy}.

\begin{table}[!t]
\caption{Bit Layout of the 60-Bit Resource Descriptor}
\label{tab:resource}
\centering
\begin{tabular}{@{}rrl@{}}
\hline
Offset & Width & Field or candidate interpretation\\
\hline
0 & 12 & SID label\\
12 & 2 & Zero\\
14 & 2 & Record ordinal within its group\\
16 & 7 & First-axis start \(a\)\\
23 & 7 & First-axis extent \(d\)\\
30 & 4 & Second-axis start \(b\)\\
34 & 4 & Second-axis extent minus one \(e\)\\
38 & 13 & Uninterpreted bits\\
51 & 1 & One\\
52 & 7 & Uninterpreted bits\\
59 & 1 & Zero\\
\hline
\end{tabular}
\end{table}

Each group is a maximal run of consecutive records sharing first-axis start $a$ and extent $d$. With groups determined from these two fields, bits~14--15 of the resource descriptors formed the sequence $0,\ldots,m-1$ in every group of $m$ records. The sequence restarted at zero in each of the 9663 groups, supporting the interpretation of bits~14--15 as a zero-based ordinal within the group. The 1355 lists contained 25,637 records. Groups of one, two, three, and four records occurred 2424, 2405, 933, and 3901 times, respectively.

\begin{figure}[!t]
\centering
\includegraphics[width=0.96\columnwidth]{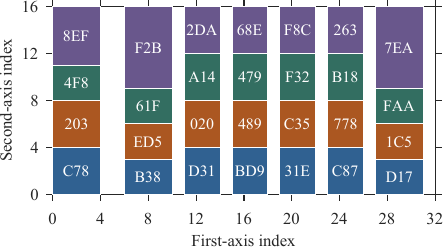}
\caption{Two-dimensional geometry of the first type-\hex{01} resource list in F0001. The 28 rectangles form seven groups in internal coordinates. Labels give hexadecimal SID values, and colors indicate the ordinal within a group.}
\label{fig:map}
\end{figure}

The low seven bits of group descriptor $j$ repeat $a_j$, and its high bit $u_j$ specifies the gap to the following group. The relation $a_{j+1}=a_j+d_j+1+u_j$ held for all 8308 adjacent-group pairs. In every group, the second-axis intervals partitioned $[0,16)$ without gaps or overlap. Fig.~\ref{fig:map} shows the decoded list in F0001. Its first record carries SID \hex{C78} and has group ordinal~0. The fields \((a,d,b,e)=(0,4,0,3)\) give the rectangle \([0,4)\times[0,4)\). The other three records in the group share the first-axis interval and cover \([4,8)\), \([8,11)\), and \([11,16)\) on the second axis.

\begin{figure}[!t]
\centering
\includegraphics[width=0.98\columnwidth]{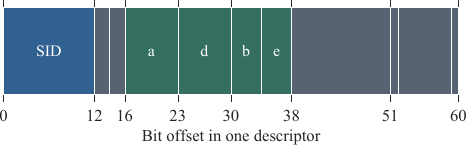}
\caption{Bit layout of a 60-bit descriptor with symbols for the decoded coordinates. The gray blocks include constant, ordinal, and uninterpreted fields detailed in Table~\ref{tab:resource}.}
\label{fig:grantaudit}
\end{figure}

\subsection{Evidence for Candidate Uplink Grants}
The resource groups identified in Section~IV-B support interpreting the first axis as time and the second as frequency. Each group contains up to four records that share the same first-axis start and extent, while their disjoint second-axis intervals together cover 16 units. This structure is consistent with the uplink burst arrangement illustrated in Fig.~2 of the SpaceX scheduler patent, which describes an option in which terminals within a burst receive allocations of equal duration and gives an example with 16 frequency-domain resource blocks \cite{scheduler}. The description accompanying Fig.~5 of the same patent further explains that the satellite can broadcast an uplink map specifying a time start, a symbol count, a frequency start, and a frequency-resource count for each terminal. Under this interpretation, those four quantities correspond to the decoded fields \(a\), \(d\), \(b\), and \(e+1\), respectively, and the 12-bit SID label is a candidate addressing field. Table~\ref{tab:patent} summarizes these correspondences.

\begin{table}[!t]
\caption{Possible Correspondences With the Patented Uplink Map}
\label{tab:patent}
\centering
\footnotesize
\begin{tabular}{@{}lp{0.43\columnwidth}@{}}
\hline
Patent grant information & Observed descriptor\\
\hline
User terminal association & 12-bit SID label\\
Starting time in the frame & First-axis start $a$\\
Number of time symbols & First-axis extent $d$\\
Starting frequency resource & Second-axis start $b$\\
Number of frequency resources & Second-axis count $e+1$\\
\hline
\end{tabular}
\end{table}

The patent also describes downlink maps with analogous time and frequency parameters \cite{scheduler}. The shared first-axis interval and the partition into up to four allocations provide the more specific correspondence to its uplink burst example. Together, the observed geometry and field correspondence support interpreting the descriptors as candidate uplink grants. Independent uplink observations are needed to establish the physical units, associate SID labels with individual terminals, and verify the scheduled transmissions.

\section{Conclusion}
The inter-header MCS-5 allocation carries a CRC-24-protected message directory, satellite state and timing information, and a fixed ground reference. The shared resource-list format in type-\hex{01} and type-\hex{03} messages yields decoded list lengths that match the inner-header length descriptors in all 1355 list-bearing messages. Their 60-bit descriptors encode labeled two-dimensional resource allocations. The observed grouping and partitioning are consistent with the uplink resource map described in a SpaceX patent. Confirming this interpretation requires uplink observations to calibrate the resource axes, associate resource labels with terminals, and verify scheduled transmissions.

\section*{Data Availability}
The data used in this study were obtained from the University of Texas at Austin's public repository for the study by Qin et al.~\cite{qin2026,qindata}.

\bibliographystyle{IEEEtran}
\bibliography{references}
\end{document}